\documentstyle{article}
\title{Estimates on Green functions of
second order differential operators with singular coefficients }
\author{ Z. Haba\\Institute of Theoretical Physics, University of Wroclaw,
\\50-204 Wroclaw, Plac Maxa Borna 9, Poland\\e-mail:zhab@ift.uni.wroc.pl}
\date{PACS:03.70.+k,04.62.+v,02.30.Jr,02.40.Ky, 02.50.Ga}
\begin{document}
\maketitle
\begin{abstract}
We investigate the Green's functions $G(x;x^{\prime}) $ of some
second order differential operators on $R^{d+1}$ with singular
coefficients depending only on one coordinate $x_{0}$. We express
the Green's functions by means of the Brownian motion. Applying
probabilistic methods we prove that when $x=(0,{\bf x})$ and
$x^{\prime}=(0,{\bf x}^{\prime})$ (here $x_{0}=0$) lie on the
singular hyperplanes then $G(0,{\bf x};0,{\bf x}^{\prime}) $ is
more regular than the Green's function of operators with regular
coefficients.
\end{abstract}
\section{ Introduction}
We discuss Green's functions of some second order differential
operators with singular coefficients appearing in quantum physics.

As a first example  consider the Lagrangian for a scalar field in
$(d+1)$-dimensions interacting with gravity
\begin{equation}
{\cal L}= g^{\mu\nu}\partial_{\mu}\phi\partial_{\nu}\phi +
(m^{2}+\xi R)\phi^{2}
\end{equation}
where $g^{\mu\nu}$ is the metric tensor and $R$ is the scalar
curvature. Such a Lagrangian with $m=0$ and the minimal coupling
$\xi=0$ appears also in the theory of structure formation
(cosmological perturbations)\cite{bardeen}. We discuss the
Euclidean version of a spatially homogeneous metric (we write
$x=(t,{\bf x})$ or $x=(x_{0},{\bf x})$ depending on whether the
first coordinate has an interpretation of time or space)
\begin{displaymath} ds^{2}=dt^{2}+g_{jk}(t)dx^{j}dx^{k}
\end{displaymath}
The Laplace-Beltrami operator resulting from the bilinear form in
eq.(1) reads
\begin{equation}
\triangle_{g}=\frac{1}{2}g^{-\frac{1}{2}}\partial_{\mu}(g^{\mu\nu}g^{\frac{1}{2}}\partial_{\nu})
\end{equation}(here $g= \det (g_{\mu\nu})$).
In cosmological models $g_{jk}\simeq t^{2\alpha}$ and
$g^{jk}\simeq t^{-2\alpha}$ when $t\rightarrow 0$ with $\alpha>0$.
Such a singular behavior can appear also in models describing
collapse phenomena in general relativity \cite{szekeres}.

As a second example we  consider quantum mechanics on a
(topologically trivial) manifold with the Hamiltonian
\begin{equation} H=-\triangle_{g}+U(x_{0})
\end{equation}
(in some global coordinates $x=(x_{0},{\bf x})$).

The  Green functions of (Euclidean) quantum scalar fields (1) with
$m=0$ and the minimal coupling $\xi=0$ are solutions
 of the equation
\begin{equation}
-\triangle_{g}G=g^{-\frac{1}{2}}\delta
\end{equation}
These Green functions are also relevant  for classical field theory
because they describe a propagation of disturbances. In quantum
mechanics (3) we are  interested in the propagator kernels
\begin{equation}
\exp(-\tau H)(x_{0},{\bf x};x_{0}^{\prime},{\bf x}^{\prime})
\end{equation}
where $\tau$ is purely imaginary.

In this paper we prove that if the coefficients of the
Laplace-Beltrami operator have a power-law singularity at a
certain point $t=t_{0}$ then the Green functions $G(t_{0}, {\bf
x};t_{0},{\bf x}^{\prime})$ are more regular than the ones of
operators with regular coefficients (for regular coefficients the
Green function can be expressed by the geodesic distance
\cite{hadamard}\cite{dewitt}). In quantum field theory these Green
functions have the meaning of expectation values of quantum fields
at equal times. In quantum mechanics  the propagator (5) will have
an anomalous behavior in $\tau$. The Green function (4) can be
obtained from the propagator (5) by means of an integration over
$\tau$.
\section{The Green's functions }
Let us change coordinates
\begin{equation}
\frac{dt}{d\eta}=\sqrt{g}
\end{equation}
The Laplace-Beltrami operator (2) takes the form
\begin{equation}
2\triangle_{g}=g^{-1}\partial^{2}_{\eta}+g^{jk}\partial_{j}\partial_{k}
\end{equation}The bilinear form in  eq.(1) determines an operator
${\cal A}$ which is of the same form as $H$ in quantum mechanics
(eq.(3))
\begin{equation} {\cal A}=-\triangle_{g}+w
\end{equation}
Here, $w=\frac{1}{2}m^{2}+\frac{1}{2}\xi R$ for the scalar field
and $w=U$ for  quantum mechanics.  The Green's function of ${\cal
A} $ is a solution of the equation
\begin{equation}
-(\partial_{\eta}^{2}+gg^{jk}\partial_{j}\partial_{k}-W)G=
2\delta(\eta-\eta^{\prime})\delta({\bf x}-{\bf x}^{\prime})
\end{equation}
where we write \begin{equation} W=gw\end{equation} Together with
eq.(9) we consider the differential equation
\begin{equation}
-\partial_{\tau}P_{\tau}={\cal A}P_{\tau}
\end{equation}
with the initial condition $P_{0}(\eta,{\bf x};\eta^{\prime},{\bf
x}^{\prime})=\delta(\eta-\eta^{\prime})\delta({\bf x}-{\bf
x}^{\prime})$. Eq.(11) defines the transition function of a
stochastic process \cite{ikeda}.

 We can
formulate the problem of solving the equation \begin{equation}
{\cal A}G=\delta \end{equation} as a problem in the Hilbert space
of square integrable functions $L^{2}(d\eta d{\bf x})$
\cite{maurin}. We assume that W is a non-negative function. The
operator ${\cal A}$ can be considered as a self-adjoint
non-negative operator in $L^{2}$ if $gg^{jk}$ and $W$ are locally
integrable functions (then we can define the Friedrichs extension
\cite{maurin} of the symmetric differential operator (8)). The
transition function $P_{\tau}$ of eq.(11) can be defined as the
integral kernel of $\exp(-\tau{\cal A})$. Then, the kernel of the
inverse
\begin{displaymath} {\cal A}^{-1}=
\int_{0}^{\infty}d\tau\exp(-\tau{\cal A}) \end{displaymath} is the
solution of eq.(9). It follows that the Fourier transform
$\tilde{G}$ of $G$ has the representation
\begin{equation} \tilde{G}(\eta,\eta^{\prime};{\bf p})
=\int_{0}^{\infty}d\tau \tilde{P}_{\tau}(\eta,\eta^{\prime},{\bf
p})
\end{equation}
where $\tilde{P}$ is a solution of the equation
 \begin{equation}
  -\partial_{\tau}\tilde{P}_{\tau}=\tilde{{\cal
  A}}\tilde{P}_{\tau}
  \end{equation}
  with the initial condition
  $\tilde{P}_{0}(\eta,\eta^{\prime},{\bf p})=\delta(\eta-\eta^{\prime})$
  ( the fundamental solution). Here
  \begin{equation}
  \tilde{{\cal
  A}}=-\frac{1}{2}\partial_{\eta}^{2}
  +\frac{1}{2}p_{j}g^{jk}(\eta)g(\eta)p_{k}+W
  \equiv-\frac{1}{2}\partial_{\eta}^{2}+V(\eta)+W(\eta)
  \end{equation}
  Eq.(14) is a Schr\"odinger-type equation with the
  Hamiltonian $\tilde{{\cal A}}$ and the potential $V+W$ where  \begin{equation}
  V(\eta)=
  \frac{1}{2}p_{j}g^{jk}(\eta)g(\eta)p_{k}\equiv
  {\bf p}\tilde{V}{\bf p}(\eta)
  \end{equation}
  If the potentials $V$ and $W$ belong to $ L_{loc}^{1}(d\eta)$ then $\tilde{{\cal A}} $ is a well-defined
  essentially self-adjoint operator in $L^{2}(d\eta)$\cite{kato}.

  We can express the kernel ${\tilde P}$ by means of the Brownian motion
  $b$ (the Feynman-Kac formula \cite{simon}; a discussion of the probabilistic representation
  for singular potentials can be found in \cite{klauder})

\begin{equation}
\begin{array}{l}
\tilde{P}_{\tau}(\eta,\eta^{\prime},{\bf p})=
 \cr E[\delta\left(
\eta^{\prime}-\eta- b\left(\tau\right)\right)
\exp\left(-\int_{0}^{\tau}
V\left(\eta+b\left(s\right)\right)ds-\int_{0}^{\tau}
W\left(\eta+b\left(s\right)\right)ds\right)]
\end{array}
\end{equation}
where $E[.]$ denotes an average over the Brownian paths. Now, the
kernel of $\exp(-\tau {\cal A})$ has the representation
\begin{equation}
\begin{array}{l}
P_{\tau}(\eta,{\bf x},\eta^{\prime},{\bf
x}^{\prime})=(2\pi)^{-d}\int d{\bf p}
         \exp\left(i{\bf p}\left({\bf x}^{\prime}-{\bf
x}\right)\right)
 \cr E[\delta\left(
\eta^{\prime}-\eta- b\left(\tau\right)\right)
\exp\left(-\int_{0}^{\tau}
V\left(\eta+b\left(s\right)\right)ds-\int_{0}^{\tau}
W\left(\eta+b\left(s\right)\right)ds\right)]
\end{array}
\end{equation}
In order to eliminate the $\delta$ function in eq.(17) it is
useful to express the expectation value over the Brownian motion
by means of an expectation value over the Brownian bridge
$\gamma$. Let $q$ be a path connecting $\eta$ with
$\eta^{\prime}$\begin{equation}
q(\frac{s}{\tau})=\eta+(\eta^{\prime}-\eta)\frac{s}{\tau}+\sqrt{\tau}\gamma(\frac{s}{\tau})
\end{equation}
where $\gamma$ is the Gaussian process on the interval $[0,1]$
(the Brownian bridge) starting from $0$ and ending in $0$ with the
covariance
\begin{displaymath}
E[\gamma(s)\gamma(s^{\prime})]=s^{\prime}(1-s)
\end{displaymath}
 for $s^{\prime}\leq s$.
Then, eq.(17) can be rewritten in the form \cite{simon}
\begin{equation}
\begin{array}{l}
\tilde{P}_{\tau}(\eta,\eta^{\prime},{\bf p})=
(2\pi\tau)^{-\frac{1}{2}}\exp(-\frac{1}{2\tau}(\eta^{\prime}-\eta)^{2})\cr
E[\exp\left(-\tau\int_{0}^{1}ds\left(
V\left(q\left(s\right)\right)+
W\left(q\left(s\right)\right)\right)\right)
\end{array}
\end{equation}

 Applying
the Jensen inequality ( see \cite{jen1}-\cite{jen2}) to the
$E[..]$ integral we obtain the inequality
\begin{equation}
\begin{array}{l}
\tilde{P}_{\tau}(\eta,\eta^{\prime},{\bf p})\cr
\geq(2\pi\tau)^{-\frac{1}{2}}\exp(-\frac{1}{2\tau}(\eta^{\prime}-\eta)^{2})\cr
\exp\left(-\tau\int_{0}^{1}ds E[V\left(q\left(s\right)\right)+
W\left(q\left(s\right)\right)]\right)\equiv \tilde{P}^{L}
\end{array}
\end{equation}
This integral is
\begin{equation}
\begin{array}{l}
\tilde{P}_{\tau}^{L}(\eta,\eta^{\prime},{\bf p})=
(2\pi\tau)^{-\frac{1}{2}}\exp(-\frac{1}{2\tau}(\eta^{\prime}-\eta)^{2})
\cr\exp\left(-\tau\int_{0}^{1}ds \int dy(2\pi
s(1-s))^{-\frac{1}{2}}\exp(-\frac{y^{2}}{2s(1-s)})(V+W)\left(\eta+s(\eta^{\prime}-\eta)+
\sqrt{\tau}y\right)\right)
\end{array}
\end{equation}
As a simple application of the inequality (21) we note that if
\begin{equation}
V+W\leq A^{\prime}{\bf p}^{2}+B^{\prime} \end{equation} then
\begin{equation}
\begin{array}{l}
\tilde{P}_{\tau}^{L}(\eta,\eta^{\prime},{\bf p})
\geq(2\pi\tau)^{-\frac{1}{2}}\exp(-\frac{1}{2\tau}(\eta^{\prime}-\eta)^{2})
 \exp\left(-\tau A^{\prime}{\bf p}^{2}-\tau
B^{\prime}\right)
\end{array}
\end{equation}
Hence, we obtain a bound from below by
 the transition function for
the $d$-dimensional Brownian motion.

 On the other hand we
may apply the Jensen inequality in the opposite direction to the
$s$-integral

\begin{equation}
\begin{array}{l}
\tilde{P}_{\tau}(\eta,\eta^{\prime},{\bf p})\cr
\leq(2\pi\tau)^{-\frac{1}{2}}\exp(-\frac{1}{2\tau}(\eta^{\prime}-\eta)^{2})
 \int_{0}^{1}ds E[
\exp\left(-\tau V\left(q\left(s\right)\right)- \tau
W\left(q\left(s\right)\right)\right)]\equiv \tilde{P}^{U}
\end{array}
\end{equation}
This integral takes the form
\begin{equation}
\begin{array}{l}
\tilde{P}_{\tau}^{U}(\eta,\eta^{\prime},{\bf p})\cr =
 (2\pi\tau)^{-\frac{1}{2}}\exp(-\frac{1}{2\tau}(\eta^{\prime}-\eta)^{2})\cr\int_{0}^{1}ds\int dy(2\pi
s(1-s))^{-\frac{1}{2}}\exp(-\frac{y^{2}}{2s(1-s)}) \exp\left(-
\tau(V+W)\left(\eta+s(\eta^{\prime}-\eta)+\sqrt{\tau}y\right)\right)
\end{array}
\end{equation}
If \begin{displaymath} V+W\geq A{\bf p}^{2}+B
\end{displaymath}
then
\begin{equation}
\begin{array}{l}
\tilde{P}_{\tau}^{U}(\eta,\eta^{\prime},{\bf p})\cr \leq
(2\pi\tau)^{-\frac{1}{2}}\exp(-\frac{1}{2\tau}(\eta^{\prime}-\eta)^{2})
 \exp\left(-\tau A{\bf p}^{2}-\tau
B\right)\end{array}
\end{equation}
Hence,  we estimate the transition function from above by the
Wiener transition function.

\section{Scale invariant metrics} We
consider in this section a power-law cosmological expansion. Such
an expansion is an exact solution of coupled Einstein equations
for a metric and for the scalar field with an exponential
self-interaction . Some consequences for a structure formation
with such  an expansion are discussed in \cite{luc}\cite{ratra}.
If $g_{jk}(t)$ has an isotropic power-law behavior then $V$ is
scale invariant. Let us assume here  that $V$ and $W$ are
nonnegative and scale invariant around
 $\eta=0$ (there is nothing special in the choice of $\eta=0$
 as a singular point, see a discussion at eq.(44))
\begin{equation} \tilde{V}^{jk}(\lambda\eta)=\lambda
^{2\nu}\tilde{V}^{jk}(\eta)
\end{equation}
and \begin{equation}
W(\lambda\eta)=\lambda^{2\sigma}W(\eta)\end{equation}
 Let us denote
$\theta=\tau^{-\frac{1}{2}}\eta$. We apply the scaling properties
of the Brownian bridge (19). Then, for $V$ of the form (28) and
$W$ (29) we obtain
\begin{equation}
\begin{array}{l}
\tilde{P}_{\tau}(\eta,\eta^{\prime},{\bf p}) =
(2\pi\tau)^{-\frac{1}{2}}\exp(-\frac{1}{2\tau}(\eta^{\prime}-\eta)^{2})
 \cr E[
\exp\left(-\tau^{1+\nu}\int_{0}^{1} {\bf p}\tilde{V}{\bf
p}\left(\theta+s(\theta^{\prime}-\theta)
+\gamma\left(s\right)\right)ds\right)\cr\exp\left(-\tau^{1+\sigma}\int_{0}^{1}
W\left(\theta+s(\theta^{\prime}-\theta)
+\gamma\left(s\right)\right)ds\right)]
\end{array}
\end{equation}
The bounds (22) and (26) become simple if $\eta=\eta^{\prime}=0$.
Then, the bound (22) reads
\begin{equation}
\begin{array}{l}
\tilde{P}_{\tau}^{L}(0,0,{\bf p})= (2\pi\tau)^{-\frac{1}{2}}
\cr\exp\Big(-\tau^{1+\nu}\int_{0}^{1}ds \int dy(2\pi
s(1-s))^{-\frac{1}{2}}V(y)\exp(-\frac{y^{2}}{2s(1-s)})\cr
-\tau^{1+\sigma}\int_{0}^{1}ds \int dy(2\pi
s(1-s))^{-\frac{1}{2}}W(y)\exp(-\frac{y^{2}}{2s(1-s)})\Big) \cr =
(2\pi\tau)^{-\frac{1}{2}}\exp\left(-\tau^{1+\nu}{\bf p}h{\bf
p}\int_{0}^{1}ds (s(1-s))^{\nu}-B\tau^{1+\sigma}\int_{0}^{1}ds
(s(1-s))^{\sigma}\right)
\end{array}
\end{equation}
where the bilinear form $h$ in ${\bf p}$ is defined
by
\begin{equation}{\bf p} h{\bf p}=(2\pi)^{-\frac{1}{2}}\int
 dy\exp(-\frac{y^{2}}{2}){\bf p}\tilde{V}(y){\bf p}
\end{equation}and the constant $B$ in eq.(31) is
\begin{equation}B=(2\pi)^{-\frac{1}{2}}\int
 dy\exp(-\frac{y^{2}}{2})W(y)
\end{equation}The integral (32) is finite if $\nu>-\frac{1}{2}$ and
 (33) is finite if $\sigma>-\frac{1}{2}$. In such a
case the lower bound (31) is non-trivial. The upper bound (26)
takes the form
\begin{equation}
\begin{array}{l}
\tilde{P}_{\tau}^{U}(0,0,{\bf p}) =
 (2\pi\tau)^{-\frac{1}{2}}\int_{0}^{1}ds\int dy(2\pi
s(1-s))^{-\frac{1}{2}}\cr \exp(-\frac{y^{2}}{2s(1-s)}) \exp\left(-
\tau^{1+\nu}V(y)-\tau^{1+\sigma}W(y)\right)
\end{array}
\end{equation}

We are interested in the Green functions (9)
 of the operator ${\cal
A}$ which according to eqs.(13) and (18) are expressed by an
$\tau$ integration upon $P_{\tau}$. As the simplest example of the
integral (13) let $V+W=A{\bf p}^{2}+B$ then performing the
$\tau$ integration upon the rhs of eq.(13) we obtain
\begin{equation}
\tilde{G}_{0}(0,0;{\bf p})=(2A{\bf p}^{2}+2B)^{-\frac{1}{2}}
\end{equation}
This is the standard behavior of equal-time Green's functions for
the quantum  free field.

In eq.(31) let us first discuss the case $W=B=0$. Then, the
integral over $\tau$ of eq.(31) gives the lower bound on the
Green's function
\begin{equation}
\tilde{G}(0,0;{\bf p})\geq K_{1}({\bf p}h{\bf p})^{-\omega}
\end{equation}
where
\begin{equation}
\omega=\frac{1}{2(1+\nu)}
\end{equation}
In order to estimate the upper bound (34) (for $W=0$) let us
assume a lower bound $\vert \tilde{V}\vert_{0}$ on $\tilde{V}$,
i.e., for ${\bf p}\neq 0$
\begin{equation}{\bf p}\tilde{V}{\bf p}\geq {\bf
p}^{2}\vert \tilde{V}\vert_{0}>0\end{equation}
 Now, we change variables  in eqs.(13) and (34) $(\tau,y)\rightarrow (\rho,u)$
 where
 \begin{displaymath}
 \rho=\tau\vert{\bf
p}\vert^{\frac{2}{1+\nu}}\vert
\tilde{V}(y)\vert_{0}^{\frac{1}{1+\nu}}
\end{displaymath}
\begin{displaymath}
u=y(s(1-s))^{-\frac{1}{2}}
\end{displaymath}
 Then, the upper bound takes the form
\begin{equation}
\begin{array}{l}
\tilde{G}(0,0;{\bf p})\leq \vert {\bf
p}\vert^{-2\omega}\int_{0}^{\infty}d\rho
 (2\pi \rho)^{-\frac{1}{2}}\cr\int_{0}^{1}ds\int
 du(2\pi)^{-\frac{1}{2}}
(s(1-s))^{-\frac{\nu}{1+\nu}}\vert
\tilde{V}(u)\vert_{0}^{-\omega} \exp(-\frac{u^{2}}{2})
\exp\left(-\rho^{1+\nu}\right)
\end{array}
\end{equation}
We can see that the integral on the rhs of eq.(39) is finite if
$-1 <\nu <\infty$.

We can summarize our results as

{\bf Theorem 1}

Assume that $W=0$ and the potential $V$ in eq.(15) is nonnegative
and scale invariant with $\nu>-\frac{1}{2}$ (eq.(28)). Then, the
operator $\tilde{{\cal A}}$ is essentially self-adjoint and the
integral kernel of $\exp(-\tau \tilde{{\cal A}})$ has the
probabilistic representation (20). The Green's function $G$ of
eq.(9) can be defined as an integral kernel of ${\cal A}^{-1}$.
Assume that the potential $V$ satisfies the lower bound (38) then
the Fourier transform $\tilde{G}(\eta,\eta^{\prime};{\bf p})$ of
$G(\eta,\eta^{\prime},{\bf x}-{\bf x}^{\prime})$ at
$\eta=\eta^{\prime}=0$ for any ${\bf p}$ satisfies the
inequalities
\begin{equation}
K_{1}( {\bf p}h{\bf p})^{-\omega}\leq \tilde{G}(0,0;{\bf p})\leq
K_{2} \vert {\bf p}\vert^{-2\omega}
\end{equation}
where $h$ is defined in eq.(32), $K_{1}$ and $K_{2}$ are some
positive constants.

 For $\nu<0$ the Fourier transform $ \tilde{G}$ is decaying  to zero faster than
 the Green function for operators with constant coefficients. As a consequence
  $G$  is
less singular than the one for operators with constant
coefficients (see eq.(43) below).

In the configuration space if $W=0$  then we can extract the
$\tau$ dependence from $V$ using its scale invariance. Then,
changing the integration variable in eq.(18) ${\bf
p}=\tau^{-\frac{1}{2}(1+\nu)}{\bf k}$ we can conclude that $P$ has
the form
\begin{equation} P_{\tau}(\eta,{\bf x},\eta^{\prime},{\bf x}^{\prime})
 ={\tau}^{-\frac{1}{2}(1+\nu)d-\frac{1}{2}}
F(\tau^{-\frac{1}{2}}\eta,\tau^{-\frac{1}{2}}\eta^{\prime},\tau^{-\frac{1}{2}(1+\nu)}
({\bf x}-{\bf x}^{\prime}))
\end{equation}
with a certain function $F$. Integration over $\tau$ with a
rescaled $\tau=r\vert {\bf x}-{\bf
x}^{\prime}\vert^{\frac{2}{1+\nu}}$ brings the Green's function at
equal time to the form

\begin{equation} G(\eta,{\bf x},\eta,{\bf x}^{\prime})
 =\vert {\bf x}-{\bf x}^{\prime}\vert^{-d+\frac{1}{1+\nu}}
f(\vert{\bf x}-{\bf x}^{\prime}\vert^{\frac{1}{1+\nu}}\eta, ({\bf
x}-{\bf x}^{\prime})\vert{\bf x}-{\bf x}^{\prime}\vert^{-1})
\end{equation}
It follows
\begin{equation} G(0,{\bf x},0,{\bf
x}^{\prime})=\vert{\bf x}-{\bf
x}^{\prime}\vert^{-d+\frac{1}{1+\nu}}f(({\bf x}-{\bf
x}^{\prime})\vert{\bf x}-{\bf x}^{\prime}\vert^{-1})
\end{equation}
We  obtain such a behavior in $\vert {\bf x}-{\bf
x}^{\prime}\vert$ if we apply the inverse Fourier transform to the
functions on both sides of the inequalities (40).

Let us note that if $V$ is singular at $\eta_{0}\neq 0$ (e.g.,
$V\simeq \vert \eta-\eta_{0}\vert^{2\nu})$ then all our results
concerning the transition functions and Green functions still hold
true but instead of setting $\eta=\eta^{\prime}=0$ we set
$\eta=\eta^{\prime}=\eta_{0}$ (this conclusion follows directly
from eq.(30)). So, e.g., the formula (43) reads\begin{equation}
G(\eta_{0},{\bf x},\eta_{0},{\bf x}^{\prime})=\vert{\bf x}-{\bf
x}^{\prime}\vert^{-d+\frac{1}{1+\nu}}f(({\bf x}-{\bf
x}^{\prime})\vert{\bf x}-{\bf x}^{\prime}\vert^{-1})
\end{equation}

We admit now $W\neq 0$

{\bf Theorem 2}

Let $W\geq 0$ be scale invariant (eq.(29)) and
$\sigma>-\frac{1}{2}$ then (under the assumptions of Theorem 1
concerning $V$) for any $\Lambda>0$ if $\vert {\bf
p}\vert>\Lambda$ then there exist positive constants $K_{1}$ and
$K_{2}$ such that the inequalities (40) hold true.

{\bf Proof}: setting $W=0$ in eq.(34) we obtain the  upper bound
(39). For the lower bound we note that the exponential in eq.(31)
is dominated by the term quadratic in the momenta. We change the
integration variable in eqs.(13) and (34)
\begin{displaymath}
\tau=r({\bf p}h{\bf p})^{-\frac{1}{1+\nu}}
\end{displaymath}
Then, we can see that for any $\Lambda>0$ if $\vert{\bf
p}\vert>\Lambda$ then there exists a constant $C$ such that in the
exponential of eq.(31) $B\tau^{1+\sigma}< Cr^{1+\sigma}$. Then
\begin{displaymath}
\begin{array}{l}
\int d\tau\tilde{P}_{\tau}^{L}(0,0,{\bf p})\geq\cr ({\bf
p}h_{2}{\bf p})^{-\omega}\int dr (2\pi
r)^{-\frac{1}{2}}\exp\left(-r^{1+\nu}\int_{0}^{1}ds
(s(1-s))^{\nu}-Cr^{1+\sigma}\int_{0}^{1}ds
(s(1-s))^{\sigma}\right)
\end{array}
\end{displaymath}
From this lower bound and from the upper bound (39) we obtain the
results of the theorem.

 If $W>0$ then the
lower bound in eq.(40) cannot be true for arbitrarily small ${\bf
p}$ because as follows from eq.(34) ($V=0$ for ${\bf p}=0$)
\begin{equation}
\begin{array}{l}
\tilde{G}(0,0,{\bf 0}) \leq\int_{0}^{\infty}d\tau
 (2\pi\tau)^{-\frac{1}{2}}\int_{0}^{1}ds\int dy(2\pi
s(1-s))^{-\frac{1}{2}}\cr \exp(-\frac{y^{2}}{2s(1-s)})
\exp\left(-\tau^{1+\sigma}W(y)\right)
 <\infty
\end{array}
\end{equation}

 If we imposed the
condition that $t\geq 0$ (which is quite artificial in the
Euclidean framework) then we would need to impose boundary
conditions at $\eta=0$ on the Brownian motion in the path integral
(17). The Dirichlet boundary conditions can easily be imposed in
the functional integration framework. We just insert the
characteristic function of the positive real axis in the path
integral (17) rejecting all the Brownian paths which leave the
positive real axis. With the Dirichlet boundary conditions  our
estimates on the upper  bound remain unchanged whereas the
estimates on the lower bound require some minor modifications.

Let us  consider  an example of a threedimensional space. By a
change of coordinates we can diagonalize the matrix $(g_{jk})$
\begin{equation}
g_{jk}=\delta_{jk}a_{j}^{2}
\end{equation}
Let $a =(a_{1}a_{2}a_{3})^{\frac{1}{3}}$ and
\begin{displaymath}
\delta_{j}=a_{j}^{-1}a^{-2}\partial_{\eta}a_{j}
\end{displaymath}
\begin{displaymath}
\delta=a^{-3}\partial_{\eta}a \end{displaymath}
\begin{displaymath}
Q=\frac{1}{18}\sum_{j<k}(\delta_{j}-\delta_{k})^{2}
\end{displaymath}
Then, in the potential $W$ of eq.(9)\cite{fulling}
\begin{equation} gR=6a^{4}(a^{-2}\partial_{\eta}\delta
+\delta^{2}+Q)
\end{equation} and
\begin{equation} m^{2}g=m^{2}a^{6}
\end{equation}
We obtain a scale invariant $V$ and $W$ if $a_{j}$ are scale
invariant. Let us consider the simplest case when all $a_{j}$ are
equal, $t\in R$ and
\begin{equation}
a(t)=\vert t\vert^{\alpha}
\end{equation}
We have
\begin{displaymath}
\eta=(1-3\alpha)^{-1}t\vert t\vert^{-3\alpha}
\end{displaymath}
Note that for $\alpha>\frac{1}{3}$ the point $t=0$ corresponds to
$\eta=-\infty$ and $t=\infty$ to $\eta=0$.

 Then
\begin{displaymath} V(y)=\kappa{\bf p}^{2}\vert
y\vert^{2\nu}\end{displaymath} where $\kappa >0$ is a certain
constant and
\begin{equation} \nu=2\alpha(1-3\alpha)^{-1}
\end{equation}For a scale invariant metric
\begin{displaymath}
W=m^{2}g(\eta)+\xi g
R=C_{1}m^{2}\vert\eta\vert^{\frac{6\alpha}{1-3\alpha}} +\xi
C_{2}\eta^{-2}
\end{displaymath}
De Sitter space can be obtained as a limit $\alpha\rightarrow
\infty$. Then, we have $V(\eta)=c{\bf
p}^{2}\vert\eta\vert^{-\frac{4}{3}}$ and $m^{2}g=c^{\prime}
\eta^{-2}$ , hence $W(\eta)=\tilde{c} \eta^{-2}$. This is a singular
perturbation which goes beyond our analysis. It can be treated by
means of the path integral methods. However, in such a case $W$
needs a regularization, then a renormalization and a subsequent
removal of the regularization \cite{klauder}. The $\eta^{-2}$ singularity
comes also from the term $gR$. Hence, the results
of this section apply only to $\xi=0$. Then, in eq.(29)
$\sigma=3\alpha(1-3\alpha)^{-1}$. $B$ in eq.(33) is finite if
$\vert\alpha\vert<\frac{1}{3}$.

In quantum mechanics $x_{0}$ is interpreted as a space variable.
The metric  takes the form ($d+1=3$) \begin{displaymath} ds^{2}=
dx_{0}^{2}+\vert x_{0}\vert^{2\alpha}(dx_{1}^{2}+dx_{2}^{2})
\end{displaymath}
Then, $\eta=(1-2\alpha)^{-1}x_{0}\vert x_{0}\vert^{-2\alpha}$. The
Hamiltonian (3) is symmetric in $L^{2}(\sqrt{g}dx)$. The change of
coordinates $x_{0}\rightarrow \eta$ associates with $H$ the
operator $\tilde{{\cal A}}=g\tilde{H}$ which is symmetric in
$L^{2}(d\eta d{\bf x})$
\begin{displaymath}
\tilde{{\cal A}}=-\partial_{\eta}^{2}+V+W
\end{displaymath}
where
\begin{equation}
V(\eta)=C_{1}{\bf p}^{2}\vert \eta\vert^{\frac{2\alpha}{1-2\alpha}}
\end{equation}
with ${\bf p}^{2}=p_{1}^{2}+p_{2}^{2}$ and
\begin{equation}
W=gU(\eta)=C_{2}\vert\eta\vert^{\frac{4\alpha}{1-2\alpha}}U(\eta)
\end{equation}
The anomalous behavior of $\tilde{P}_{\tau}$ has as a consequence

{\bf Corollary 3}

Let $\tilde{P}_{\tau}(\eta,\eta^{\prime},{\bf p}) $ be the
fundamental solution of eq.(14) with $W=0$ and $V$ defined in eq.(51). If
$\nu=\frac{\alpha}{1-2\alpha}>-\frac{1}{2}$ then for any $\tau\geq
0$

\begin{equation}
\int d{\bf x} P_{\tau}(0,{\bf x},0,{\bf x}^{\prime})\vert {\bf x}
-{\bf x}^{\prime}\vert^{2}=(-\triangle_{\bf
p})\tilde{P}_{\tau}(0,0,{\bf p})_{\vert {\bf p}=0}=B_{1}
\tau^{\frac{1}{2}+\nu}
\end{equation}
and
\begin{equation}
\int d{\bf x} d\eta^{\prime}P_{\tau}(0,{\bf x},\eta^{\prime},{\bf
x}^{\prime})\vert {\bf x} -{\bf x}^{\prime}\vert^{2}=\int
d\eta^{\prime}(-\triangle_{\bf
p})\tilde{P}_{\tau}(0,\eta^{\prime},{\bf p})_{\vert {\bf
p}=0}=B_{2} \tau^{1+\nu}
\end{equation}
If $W(\eta)\geq 0$ defined in eq.(52) belongs to $
L_{loc}^{1}(d\eta)$ then instead of the equalities in
eqs.(53)-(54) we have  bounds from above by
$B_{1}\tau^{\frac{1}{2}+\nu} $ in eq.(53) and $B_{2}\tau^{1+\nu}$
in eq.(54).

 {\bf Proof}:we prove eq.(54) (eq.(53) is simpler and proved in a similar
 way).
  Let us
calculate
\begin{equation}
\begin{array}{l}
(-\triangle_{\bf p})\int
d\eta^{\prime}\tilde{P}_{\tau}(0,\eta^{\prime},{\bf p})_{\vert
{\bf p}=0} =\int
d\eta^{\prime}(2\pi\tau)^{-\frac{1}{2}}\exp(-\frac{1}{2\tau}(\eta^{\prime})^{2})
 \cr E[\tau^{1+\nu}\int_{0}^{1} Tr\tilde{V}\left(s\tau^{-\frac{1}{2}}\eta^{\prime}
+\gamma\left(s\right)\right)ds]= B_{2}\tau^{1+\nu}
\end{array}
\end{equation}
If $W\geq 0$ then instead of the expectation value (55) we have
\begin{displaymath}
 \begin{array}{l}
 \int
d\eta^{\prime}(2\pi\tau)^{-\frac{1}{2}}\exp(-\frac{1}{2\tau}(\eta^{\prime})^{2})
 \cr E\left[\tau^{1+\nu}\int_{0}^{1} Tr\tilde{V}\left(s\tau^{-\frac{1}{2}}\eta^{\prime}
+\gamma\left(s\right)\right)ds\exp\left(-\int_{0}^{\tau}W\left(s\tau^{-1}\eta^{\prime}
+\sqrt{\tau}\gamma\left(\frac{s}{\tau}\right)\right)ds\right)\right]\leq
B_{2}\tau^{1+\nu}
\end{array}
\end{displaymath}
where the inequality follows from $W\geq 0$.

 Corollary 3  means
that if $\nu<0$ then the sample paths of diffusions generated by
operators with singular coefficients have worse continuity
properties than the Brownian paths (for Brownian paths see
\cite{simon})).\section{More general metrics}

 We study  the lower bound on $G$ following from eq.(22)
 \begin{displaymath}
\begin{array}{l}
\tilde{G}^{L}(0,0,{\bf p})=
\int_{0}^{\infty}d\tau(2\pi\tau)^{-\frac{1}{2}}
\cr\exp\left(-\tau\int_{0}^{1}ds \int dy(2\pi
s(1-s))^{-\frac{1}{2}}\exp(-\frac{y^{2}}{2s(1-s)}){\cal V}\left(
\sqrt{\tau}y\right)\right)
\end{array}
\end{displaymath}
and  the upper bound following from eq.(26)

\begin{displaymath}
\begin{array}{l}
\tilde{G}^{U}(0,0,{\bf p})\cr =
 \int_{0}^{\infty}d\tau(2\pi\tau)^{-\frac{1}{2}}\cr\int_{0}^{1}ds\int dy(2\pi
s(1-s))^{-\frac{1}{2}}\exp(-\frac{y^{2}}{2s(1-s)}) \exp\left(-
\tau{\cal V}\left(\sqrt{\tau}y\right)\right)
\end{array}
\end{displaymath}
for more general ${\cal V}$

A generalization of Theorem 1 reads

{\bf Theorem 4}

Let us consider ${\cal V}={\bf p}\tilde{{\cal V}}{\bf p}$ which is
not scale invariant but of the form \begin{equation} \tilde{{\cal
V}}(\eta)=\tilde{V}(\eta)f(\eta)+l(\eta)
\end{equation}
where $\tilde{V}$ is a matrix scale invariant function (28)
satisfying the conditions of Theorem 1 with $-\frac{1}{2}<\nu<0$,
$f$ is a bounded function with a strictly positive lower bound,
$l$ is a nonnegative bounded matrix function. Assume in addition
that
\begin{equation}
\int dy\exp(-\frac{y^{2}}{2})f(y)\tilde{V}(y)\geq c I>0
\end{equation}
where $c$ is a positive number. Under our assumptions (56)-(57)
for any $\Lambda>0$ if $\vert{\bf p}\vert
>\Lambda$ then there exist a positively definite bilinear form $h_{2}$
  and constants $K_{1}$ and $K_{2}$ such that

\begin{equation}
K_{1}( {\bf p}h_{2}{\bf p})^{-\omega}\leq \tilde{G}(0,0;{\bf
p})\leq K_{2} \vert {\bf p}\vert^{-2\omega}
\end{equation}
If $\nu\geq 0$ for $\tilde{V}$ in eq.(56) then  for $\vert{\bf
p}\vert>\Lambda$  the inequalities (58) hold true with
$\omega=\frac{1}{2}$.

 {\bf Proof}:
   our assumptions (56) on $\tilde{{\cal V}}$ mean that it satisfies the inequalities
\begin{equation}
\tau^{\nu}{\bf p}\tilde{V}_{1}(y){\bf p}+{\bf p}l_{1}{\bf p}\leq
{\cal V}(\sqrt{\tau}y)\leq\tau^{\nu}{\bf p}\tilde{V}_{2}(y){\bf
p}+{\bf p}l_{2}{\bf p}
\end{equation}
with certain matrix functions $\tilde{V}_{1}$ and $\tilde{V}_{2}$
independent of $\tau$ and bilinear forms $l_{1}$ and $l_{2}$
(independent of $y$). It follows that the integral of $\tilde{
P}_{\tau}$ satisfies the bounds
\begin{equation}
\int d\tau\tilde{P}_{\tau}^{L_{2}}\exp(-\tau {\bf p}l_{2}{\bf
p})\leq \int d\tau\tilde{P}_{\tau}\leq \int
d\tau\tilde{P}_{\tau}^{U_{1}}\exp(-\tau{\bf p}l_{1}{\bf p})
\end{equation}
where in the lower bound  $\tilde{P}^{L_{2}}$  the potential
$V_{2}$ from the rhs of eq.(59) is applied and in
$\tilde{P}^{U_{1}}$ the one from the lhs of eq.(59).The integral
(57) defines $h$ of eq.(32) (and the $h_{2}$ from the upper bound
(59)). Let us change the integration variable $\tau = r ({\bf
p}h_{2} {\bf p})^{-\frac{1}{1+\nu}}$ on the lhs of eq. (60) and
$\tau = \rho\vert{\bf
p}\vert^{-4\omega}\vert\tilde{V}_{1}(y)\vert_{0}^{-2\omega}$ on
the rhs. Then, the lower and upper bounds read (from eqs.(31),(34)
and (38))\begin{equation}
\begin{array}{l}
({\bf p}h_{2}{\bf p})^{-\omega}\int_{0}^{\infty}dr(2\pi
r)^{-\frac{1}{2}}\exp(-r^{1+\nu}-r({\bf p}h_{2}{\bf p})^{-2\omega}
{\bf p}l_{2}{\bf p})\cr \leq\tilde{G}(0,0,{\bf p}) \leq \vert{\bf
p}\vert^{-2\omega} \int_{0}^{\infty}d\rho
 (2\pi \rho)^{-\frac{1}{2}}\cr\int_{0}^{1}ds\int
 du(2\pi)^{-\frac{1}{2}}
(s(1-s))^{-\frac{\nu}{1+\nu}}\vert
\tilde{V}_{1}(u)\vert_{0}^{-\omega}
 \exp(-\frac{u^{2}}{2})
\cr
\exp\left(-\rho^{1+\nu}-\rho \vert{\bf p}\vert^{-4\omega}\vert \tilde {V}_{1}(u\sqrt{s(1-s)})\vert_{0}^{-2\omega}
{\bf p}l_{1}{\bf p}\right)
\end{array}
\end{equation}
The condition (57) implies that the bilinear form $h_{2}$ is
strictly positive. Hence, there exists a constant $K$ such that
\begin{displaymath}K {\bf p}h_{2}{\bf p}\geq {\bf p}l_{2}{\bf
p}\end{displaymath}
  Then, for
$-\frac{1}{2}<\nu<0$ and $\vert {\bf p}\vert>\Lambda$ there exists
$c_{1}$ such that
\begin{displaymath}r({\bf p}h_{2}{\bf p})^{-\frac{1}{1+\nu}} {\bf
p}l_{2}{\bf p}<rc_{1}
\end{displaymath}in the exponential  on the lhs of eq.(61).
The $l_{1}$ term can be set zero for the upper
bound. In such a case for each $\Lambda>0$ there exist constants
$c_{1}$ and $c_{2}$ such that if $\vert {\bf p}\vert
>\Lambda$ then the inequalities (61) take the form

\begin{equation}
\begin{array}{l}
({\bf p}h_{2}{\bf p})^{-\omega}\int_{0}^{\infty}dr(2\pi
r)^{-\frac{1}{2}}\exp(-r^{1+\nu}-rc_{1})\cr \leq\tilde{G}(0,0,{\bf
p})  \leq \vert{\bf p}\vert^{-2\omega} \int_{0}^{\infty}d\rho
 (2\pi \rho)^{-\frac{1}{2}}\cr\int_{0}^{1}ds\int
 du(2\pi)^{-\frac{1}{2}}
(s(1-s))^{-\frac{\nu}{1+\nu}}\vert
\tilde{V}_{1}(u)\vert_{0}^{-\omega}
 \exp(-\frac{u^{2}}{2})
\exp\left(-\rho^{1+\nu}\right)
\end{array}
\end{equation}
The inequalities (62) coincide with (58) because under our
assumptions the integrals in eq.(62) are finite. The last
statement of Theorem 4 follows from the inequalities (60) because
the behavior for large ${\bf p}$ follows from the behavior of
$\tilde{P}_{\tau}$ for a small $\tau$. If $\nu>0$ then in eq.(60)
$\tau^{1+\nu}<A\tau$ for any $A$ and a sufficiently small $\tau$.
Hence, we obtain the same  behavior of $\tilde{G}$ for large
momenta as in the case $\tilde{V}=1$.

We would like to note that the restrictive form (56) of ${\cal V}$
is not necessary. As an example we could consider ${\cal V}$ which
has singularities at several points, e.g.
\begin{equation}
{\cal V}(\eta)={\bf p}^{2}(a_{0}\vert
\eta-\eta_{0}\vert^{2\nu_{0}}+\kappa\vert \eta\vert^{2\nu})
\end{equation}
with $\vert\nu_{0}\vert<\vert \nu\vert$ (only negative indices are
non-trivial). An application of the lower and upper bounds (31)
and (34) to the potential (63) leads to the conclusion that
after an integration upon $\tau$ the
inequalities (40) hold true for $\vert {\bf p}\vert>\Lambda$.
Hence, the leading singularity $\nu$ determines the behavior at
large momenta.

 \section{Discussion and summary }

As we pointed out in the Introduction  our results concerning the
Green functions can find an application to quantum field theory in
an expanding universe. The stronger damping in momenta (eq.(40))
in the inflationary models ($\alpha>1$) at $\eta=\eta^{\prime}=0$
indicates that it would be promising to start quantization at this
time ($\eta=0$ corresponds to $t=\infty$ in cosmological models
with $\alpha>\frac{1}{3}$). The exponential inflation can be
obtained as a limit $\alpha\rightarrow \infty$ which corresponds
to $\nu=-\frac{2}{3}$. This limit is beyond our rigorous approach
but it could be treated by means of more sophisticated methods of
ref.\cite{klauder}. By a formal scaling argument we obtain again
the behavior (40) which in inflationary cosmological models is
known as the Harrison-Zeldovich spectrum of scalar fluctuations
\cite{bardeen}\cite{wise}. The Green functions can be applied in
order to derive a solution of Einstein equations linearized around
the homogeneous background \cite{luc}\cite{ratra}. In such a case
in addition to the scalar Green function the tensor Green function
must be studied as well. Further  consequences of our estimates
concerning the spectrum of $\tilde{G}$ for the complete theory
still need to be explored. For this purpose a detailed  dependence
of the Green function on $\eta$ and $\eta^{\prime}$ would be
useful. It is much harder to derive such estimates than the ones
for the  time zero case. In the Appendix we investigate the upper
bound $G^{U}$ for general $\eta$. In particular, calculations
performed there suggest that it is only the behavior of ${\cal
V}(y)$ for small $y$ which is relevant for Theorem 4 and that the
upper bound is valid for all $\nu>-1+\frac{1}{d}$ . For the lower
bound $G^{L}$ we can also obtain an integral representation.
However, it is quite complicated.

 Another motivation for a study of the $(d+1)$-dimensional Green functions
 comes from the problem of a dimensional reduction of quantum fields
 defined on a brane \cite{dvali}.
In such a case we restrict ourselves to a $d$-dimensional
submanifold  imposing the condition $\eta=\eta^{\prime}=0$. We
have proved here that if the metric has a power-law behavior then
the Green functions of the restricted quantum field theory are
decaying faster in the momentum space than the standard $\vert{\bf
p}\vert^{-1}$. In particular, for $\nu=-\frac{1}{2}$ we obtain the
propagator $\vert {\bf p}\vert^{-2}$ in $d$-dimensions which is
the same as the one of the Euclidean massless free field.

\section{Appendix}
We calculate the upper bound for the Green's function $\tilde{G} $
following form eqs.(26) and (13) in more detail. Set
$u=\sqrt{\tau}y$ and let us perform the integration upon $\tau$ in
eq.(26) with $W=0$. Then
\begin{equation}
\begin{array}{l}
\tilde{G}^{U}(\eta,\eta^{\prime},{\bf p}) =
 2(2\pi)^{-\frac{1}{2}}\int_{0}^{1}ds\int du(2\pi
s(1-s))^{-\frac{1}{2}}K_{0}\left(\sqrt{2M{\bf p}\tilde{V}{\bf
p}\left(u\right)}\right)
\end{array}
\end{equation}
where \begin{equation}
M=(\eta^{\prime}-\eta)^{2}+\left(s\left(1-s\right)\right)^{-1}\left(u-\eta
-s\left(\eta^{\prime}-\eta\right)\right)^{2}
\end{equation}
and $K_{\rho}$ is the modified Bessel function of order $\rho$
\cite{abram}.

 The integral is simpler if
$\eta=\eta^{\prime}$\begin{equation}
\begin{array}{l}
\tilde{G}^{U}(\eta,\eta,{\bf p}) =
 \pi^{-1}\int_{0}^{1}ds\int duK_{0}\left(N\sqrt{2{\bf p}\tilde{V}{\bf
p}\left(u\right)}\right)
\end{array}
\end{equation}
where \begin{equation}
N=\left(s\left(1-s\right)\right)^{-\frac{1}{2}}\eta-u
\end{equation}If $\eta^{\prime}=\eta=0$ then the integral (64) further simplifies. If
additionally $V$ is scale invariant then we can calculate it
exactly as in sec.3.

For arbitrary $\eta$ and $\eta ^{\prime}$ the behavior of
$\tilde{G}^{U}$ is much more complicated because the decay of
$\tilde{G}^{U}$ substantially depends on $\eta$. In the simplest
case when $\tilde{V}=\frac{1}{2}$
\begin{equation}
\tilde{G}(\eta,\eta^{\prime},{\bf p})=\vert{\bf
p}\vert^{-1}\exp(-\vert{\bf p}\vert \vert \eta^{\prime}-\eta\vert)
\end{equation}
Let us consider an arbitrary $V$ and assume that it behaves as
\begin{equation} V(\lambda y)= \lambda^{2\nu} B_{\lambda}(y)
\end{equation}
when $\lambda\rightarrow 0$ with a certain $B_{\lambda}$ which as
a function of $\lambda$ is bounded from above and from
below,i.e.,$C_{2}(y)\geq B_{\lambda}(y)\geq C_{1}(y)>0$ for a
small  $\lambda$. Let us change variables in eq.(64)
\begin{equation}
u=\vert {\bf p}\vert^{-\frac{1}{1+\nu}}y
\end{equation}
Assume that $\vert{\bf p}\vert\rightarrow \infty$ ,
$\eta\rightarrow 0$ and $\eta^{\prime}\rightarrow 0$ in such a way
that $\theta = \vert {\bf p}\vert^{\frac{1}{1+\nu}}\eta$ and
$\theta^{\prime} = \vert {\bf
p}\vert^{\frac{1}{1+\nu}}\eta^{\prime}$ remain finite. In such a
case from eq.(64) we can conclude that
\begin{equation}
\tilde{G}(\vert {\bf p}\vert^{-\frac{1}{1+\nu}}\theta,\vert {\bf
p}\vert^{-\frac{1}{1+\nu}}\theta^{\prime},{\bf p}) \simeq \vert
{\bf p}\vert^{-2\omega}
\end{equation}
for large ${\bf p}$ in agreement with eq.(40) and eq.(68) ($\nu=0$
for a constant $\tilde{V}$).

 Eqs.(64)-(67) give an integral
representation of the upper bound which is expected to approximate
the exact Green function $\tilde{G}$ for large ${\bf p}$. We
suppose that the Fourier transform $G^{U}$ of $\tilde{G}^{U}$ is a
reliable approximation to $G$ at short distances. After the
Fourier transform of eq.(26) with $W=0$ we can calculate the $\tau$ integral
in eq.(13) exactly. We obtain
\begin{equation}
\begin{array}{l}
G^{U}(\eta,{\bf x},\eta^{\prime},{\bf x}^{\prime})=
(2\pi)^{-\frac{d+2}{2}}\int_{0}^{1}ds\left(s\left(1-s\right)\right)^{-\frac{1}{2}}\int
dy \left(\det \tilde{V}\left(y\right)\right)^{-\frac{1}{2}} \cr
\left(\left({\bf x}-{\bf
x}^{\prime}\right)\tilde{V}^{-1}\left(y\right)\left({\bf x}-{\bf
x}^{\prime}\right)+\left(\eta-\eta^{\prime}\right)^{2}+\left(s\left(1-s\right)\right)^{-1}
\left(y-\eta-s\left(\eta^{\prime}-\eta\right)\right)^{2}\right)^{-\frac{d}{2}}
\end{array}
\end{equation}
If $\tilde{V}^{ij}(y)=\delta^{ij}v(y)$ then the formula (72) can
be expressed in a simpler form\begin{equation}
\begin{array}{l}
G^{U}(\eta,{\bf x},\eta^{\prime},{\bf x}^{\prime})=
(2\pi)^{-\frac{d+2}{2}}\int_{0}^{1}ds\left(s\left(1-s\right)\right)^{-\frac{1}{2}}\int
dy \cr\left(\vert {\bf x}-{\bf
x}^{\prime}\vert^{2}+v(y)\left(\eta-\eta^{\prime}\right)^{2}+\left(s\left(1-s\right)\right)^{-1}
v(y)\left(y-\eta-s\left(\eta^{\prime}-\eta\right)\right)^{2}\right)^{-\frac{d}{2}}
\end{array}
\end{equation}If $v=1$ then eq.(73) gives
\begin{equation}
(2\pi)^{-\frac{d+1}{2}}\left(\vert {\bf x}-{\bf
x}^{\prime}\vert^{2}+\left(\eta-\eta^{\prime}\right)^{2}\right)^{-\frac{d-1}{2}}
\end{equation}
as it should.

 The integrals (72)-(73) suggest some generalizations
of the theorems proved in the main part. First, assume that
$V(y)\simeq \vert y\vert^{2\rho} $ for a large $\vert y\vert$ then
the integrals (72)-(73) are finite (for $\vert {\bf x}-{\bf
x}^{\prime}\vert\neq 0$) if $\rho>-1+\frac{1}{d}$. Next, it can be
shown from eq.(73) that if $v(y)\simeq\vert y\vert^{2\nu}$ for
$y\rightarrow 0$ and $\eta=\eta^{\prime}=0$ then
\begin{displaymath} G^{U}(0,{\bf x},0,{\bf x}^{\prime})\simeq \vert
{\bf x}-{\bf
x}^{\prime}\vert^{-d+\frac{1}{1+\nu}}\end{displaymath} for ${\bf
x}-{\bf x}^{\prime}\rightarrow 0$. The derivation of the result
(44) based on eq.(73) suggests that for Theorem 4 only the
behavior of $V(y)$ for a small $y$ is relevant (assuming the
integral (73) is finite ).

For general $v$ and arbitrary  ${\bf x}$,${\bf x}^{\prime}$,$\eta$
and $\eta^{\prime}$ it is harder to obtain usable estimates. Let
us mention some special cases. It follows directly from eq.(41)
that\begin{displaymath} G(\eta,{\bf x},0,{\bf
x})\simeq\vert\eta\vert^{-d(1+\nu)+1}
\end{displaymath}
and
\begin{displaymath} G(0,{\bf
x},\eta^{\prime},{\bf x
})\simeq\vert\eta^{\prime}\vert^{-d(1+\nu)+1}
\end{displaymath}
whereas from eq.(73) we obtain that if $v(\eta)\neq 0$
then\begin{equation} G^{U}(\eta,{\bf x},\eta,{\bf
x}^{\prime})\simeq v(\eta)^{-\frac{1}{2}} \vert {\bf x}-{\bf
x}^{\prime}\vert^{-d+1}\end{equation} when ${\bf x}\rightarrow
{\bf x}^{\prime}$.

If $V(y)\geq c\vert y\vert^{2\rho}$ with $c>0$ and
$\rho>-1+\frac{1}{d}$ for large $y$ then changing the integration
variable $y=\vert {\bf x}-{\bf
x}^{\prime}\vert^{\frac{1}{1+\rho}}z$ we can show
 that for any $\eta$ and $\eta^{\prime}$ there exists $A$ such that
 if $\vert {\bf x}-{\bf x}^{\prime}\vert \geq A$ then
\begin{equation}
G^{U}(\eta,{\bf x},\eta^{\prime},{\bf x}^{\prime})\leq K\vert {\bf
x}-{\bf x}^{\prime}\vert^{-d+\frac{1}{1+\rho}}
\end{equation}
When $\rho>0$ then eq.(76) gives a non-trivial estimate saying
that the Green's function has a stronger decay for large distances
than the one for operators with constant coefficients. However,
such a decay at large distances will be changed by most
perturbations $W$ whereas the behavior for short distances is
remarkably stable with respect to perturbations.

\end{document}